\documentclass[journal, a4paper]{IEEEtran}

\begin{document}

\title{Fast and compact time-resolved spectroscopy enabled by Quantum Walk Combs}

\author{\textbf{Ina Heckelmann$^{1,*}$, Davide Pinto$^{1}$, Uwe Schmitt$^{2}$, Mattias Beck$^{1}$, Jérôme Faist$^{1}$} \\
1. Institute of Quantum Electronics, Department of Physics, ETH Zürich, 8093 Zürich, Switzerland \\
2. Scientific IT Services, ETH Zürich, 8093 Zürich, Switzerland \\
* Corresponding author. Email: iheckelmann@ethz.ch}
\maketitle

\begin{abstract} 
    Optical frequency combs paved the way for fast and compact multi-component optical chemical analysis due to their broadband spectra. Dual-comb spectrometers embody this technology, but their design requires a pair of matched combs, adding complexity to the system. In this study, we propose and implement a scheme for a rapid, compact spectroscopic analyser that operates without moving components, leveraging the tunability and speed of a single quantum walk comb laser. Previously, these combs have been shown to deliver stable, tunable and broadband lasing. Here, a single quantum cascade laser-based comb emitting within the significant molecular fingerprint region of the mid-infrared spectrum was employed in a non-interferometric setup for targeted and non-targeted analysis of various organic solvent vapours. With a time resolution as small as \qty{10}{\micro\second} and a high dynamic range reaching three orders of magnitude in concentration, this approach is suitable for the real-time analysis of chemical kinetics.
\end{abstract}

\section{Introduction}

\begin{figure*}[!b]
    \centering
    \includegraphics[width=1\textwidth]{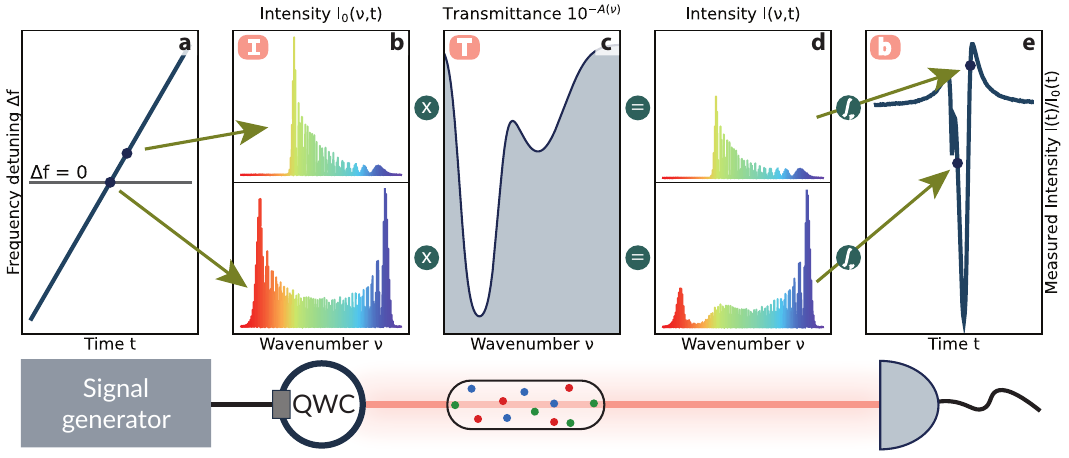}
    \caption{\textbf{Schematic of spectroscopy enabled by a tunable comb source.} The modulation frequency is chirped linearly by a signal generator \textbf{(a)}, leading to variable optical outputs of the Quantum Walk Comb source \textbf{(b)}. After the laser output passes the sample gas with some characteristic transmittance (\textbf{(c,d)}, the spectrally integrated power is measured as a function of time by a photodetector \textbf{(e)}.}
    \label{fig: overview}
\end{figure*}   

Conducting high-speed wavelength-resolved spectroscopy in the mid-infrared regime facilitates the real-time, continuous observation of rapid, non-repetitive, or infrequent phenomena, proving advantageous in areas such as kinetic studies within combustion chemistry \cite{bjork_direct_2016,pinkowski_dual-comb_2020}, free radical kinetics \cite{fleisher_mid-infrared_2014}, and protein analysis \cite{klocke_single-shot_2018,norahan_microsecond-resolved_2021}. 

In principle, the speed for taking absorption measurements without spectral resolution is constrained solely by the bandwidth of the detector, allowing potential temporal resolution in the order of \qty{10}{\pico\second} with quantum-well infrared detectors \cite{grant_room-temperature_2006}. Nonetheless, in practical settings, spectral resolution is vital for accurately identifying samples by their specific absorption properties or for concurrently analyzing concentrations of various species. To attain this, a number of strategies have been adopted to alter either the source or the instrumentation setup. A notable modification to the laser source includes the coupling of the laser chip to an external grating, which is mechanically displaced for the selection of a single longitudinal mode. This mode can be swiftly tuned over an extensive bandwidth, enabling full spectral acquisition in approximately \qty{20}{\micro\second} \cite{lyakh_external_2015,sun_detection_2018}. External cavity (EC) sources, while highly versatile, are, however, sensitive to the alignment of the laser chip to the grating. For instance, mechanical instabilities of the latter can lead to mode-hopping of the laser frequency, resulting in scan-to-scan reproducibility errors \cite{brumfield_characterization_2015, schwaighofer_external_2016}. State-of-the-art EC-Quantum Cascade Lasers can cover more than \qty{500}{\per\cm} in a multichip design, but remain bulky and expensive.

Another thoroughly investigated source modification involves using two suitably matched frequency combs in a dual-comb spectrometer, enabling continuous time-resolved measurements with temporal resolutions reaching approximately \qty{10}{\nano\second}, dependent on the frequency difference between the combs' repetition rates \cite{coddington_dual-comb_2016,long_nanosecond_2024}. Yet, to enhance the signal-to-noise ratio, data typically undergo integration over several microseconds, thus lowering practical acquisition speeds \cite{pinkowski_dual-comb_2020}. 

An example of adaptation at the instrument level involves the use of time-resolved Fourier transform infrared (FTIR) spectrometers with a broadband source. While stroboscopic techniques can approach acquisition speeds close to the detector limitations for cyclic and repetitive events \cite{palmer_breaking_1999}, they fall short in offering real-time monitoring. Dispersive spectrometers employing detector arrays, conversely, are currently restricted to acquisition rates near \qty{8}{\milli\second}, owing to the slow refresh rates of existing charge-coupled devices \cite{rockmore_vecsel-based_2020}. Ultimately, although ongoing efforts aim to fulfil the need for fast, real-time, and broadband observation of mid-infrared absorbers, present techniques are constrained by speed, as observed with dispersive spectrometers or FTIRs, or necessitate intricate setups like the exceptionally fast but experimentally demanding dual-comb spectroscopy.


In this study, we leverage the fast tunability of the recently demonstrated Quantum Walk Comb over a broad bandwidth range for direct, rapid, and compact absorption spectroscopy. Due to the fast \qty{10}{\micro\second} acquisition time of the demonstrated technique, it is suitable for fast continuous absorption measurements and real-time process monitoring, using a simple optical setup consisting only of a frequency-modulated Quantum Walk Comb (QWC) and a sufficiently fast detector. A \qty{97}{\per\cm} bandwidth transmittance spectrum is acquired by sweeping the modulation frequency through the cavity resonance (Fig. \ref{fig: overview}a), thus causing the comb to react to this electric input by rapidly redistributing its intensity across the spectrum (Fig. \ref{fig: overview}b). After this laser output has interrogated the gas cell (Fig. \ref{fig: overview}c,d), the spectrally integrated intensity is measured on a fast detector (Fig. \ref{fig: overview}e). As the response of QWCs to varying modulation frequencies is predictable and reproducible \cite{dikopoltsev_collective_2025, dikopoltsev_theory_2025}, this behaviour is used to probe solvent vapours of acetone and 2-butanone with a straightforward, non-dispersive optical setup consisting solely of the driven QWC, the gas cell, and detector (Fig. \ref{fig: overview}, bottom). 

\section{QWCs as rapidly tunable broadband sources}

\begin{figure}[ht]
    \centering
    \includegraphics[width=0.45\textwidth]{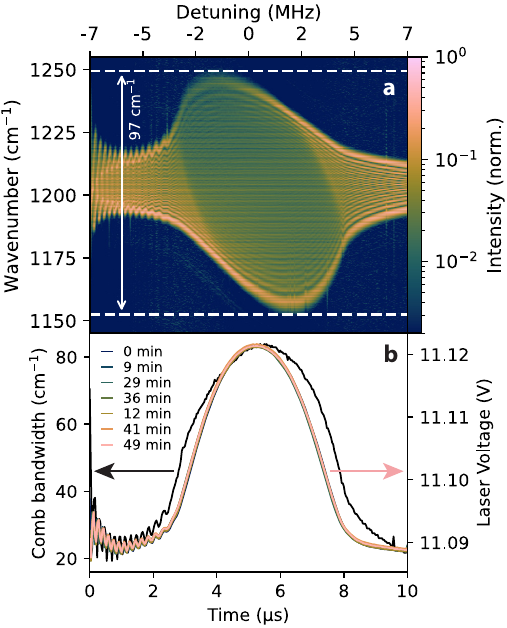}
    \caption{\textbf{Fast and reproducible tuning of a QWC.} \textbf{(a)} The comb covers a total bandwidth of \qty{97}{\per\cm} as the injection frequency is swept by \qty{14}{MHz} around its resonance within \qty{10}{\micro\second}. \textbf{(b)} Both the comb bandwidth (black, left) as well as the laser voltage (right) increase continuously as the resonance is approached, with characteristic Bloch oscillations seen at early times up until \qty{3}{\micro\second}. The voltage trace of the laser is stable over the measurement time of \qty{49}{\minute}.}
    \label{fig: comb tuning}
\end{figure}

Quantum walk combs are continuous-wave frequency modulated combs that arise in actively phase modulated ring lasers with sufficiently fast gain recovery times \cite{heckelmann_quantum_2023}. In the present study, an InGaAs/InAlAs quantum cascade laser (QCL), known for its fast upper-state lifetime of the order of \si{\pico \second} \cite{faist_quantum_1994}, is fabricated through an inverted buried heterostructure process to create a ring cavity \cite{beck_continuous_2002}. A direct current is applied to achieve lasing, complemented by a radiofrequency current for active phase modulation around the cavity resonance at \qty{14.45}{\giga\hertz}. This modulation prompts a transition from an initial free-running single-mode lasing state to a flat-top, highly stable frequency comb. While the spectra exhibit broadband and symmetrical characteristics when on resonance, they narrow upon detuning the modulation from resonance, resulting in the dominance of either the blue- or red-shifted lobe of the spectra (Fig. \ref{fig: overview}b, Fig. \ref{fig: comb tuning}a).

In this study, the modulation frequency is swept through the cavity resonance by employing a triangular frequency modulated (FM) injection with a \qty{10}{\micro\second} upward ramp, a \qty{200}{\nano\second} downward ramp and \qty{7}{\mega\hertz} deviation using a microwave signal generator (R\&S, SMF100A). As a result, the QWC expands to its full \qty{83.5}{\per \cm} bandwidth once per FM period. 
For spectral characterisation, the comb spectra are recorded with variable time delay using a boxcar averager (Zürich Instrument, UHFLI) and FTIR spectrometer (Bruker, Vertex80) during the \qty{10}{\micro\second} upwards ramp of the FM sweep (see \cite{heckelmann_quantum_2023} for experimental setup). The period of the FM sweep is ultimately limited by the speed at which the laser reacts to the change in modulation \cite{heckelmann_quantum_2023}, benchmarked by an expansion time of \qty{170}{\nano\second}, in line with values seen in literature \cite{heckelmann_quantum_2023}. If the FM period is reduced below \qty{10}{\micro\second}, Bloch oscillations seen for detuned injection at early times up to \qty{3}{\micro \second} become detrimental \cite{dikopoltsev_collective_2025}, and the maximum bandwidth is reduced. 

The capability of the QWC to respond predictably and stably to rapid changes of its driving conditions is facilitated by the fast gain recovery time of the QCL active region. By effectively suppressing any intensity fluctuations within the cavity, continuous-wave lasing with constant output intensity is maintained, resulting in the laser's light exhibiting properties akin to a liquid \cite{carusotto_quantum_2013, dikopoltsev_collective_2025, dikopoltsev_flows_2025, dikopoltsev_theory_2025}. The local intensity serves as the photon density, with strong photon-photon interactions being mediated through the fast gain nonlinearity. Consequently, the same mechanism responsible for the initial formation of QWCs also supports the fast laminar flows that occur following a quench in the laser's modulation, thereby enabling the execution of frequency modulations within the microsecond timescale without compromising coherence or reproducibility.

To experimentally assess the stability of the laser and the consequent reproducibility of the comb expansion across various FM sweeps, the laser voltage is periodically recorded using an oscilloscope (Teledyne LeCroy, HDO6104). Due to strong photon-driven transport in QCLs \cite{faist_quantum_2018}, changes in intracavity intensity lead to noticable changes in the laser's differential resistance. Consequently, the reduction in output power for resonantly driven mode-locked lasers is accompanied by an increase in differential resistance \cite{kuizenga_fm_1970}, thus establishing a correlation between the comb bandwidth and laser voltage (Fig. \ref{fig: comb tuning}b). The laser voltage is stable with average relative drifts of less than \qty{60}{ppm} (\qty{700}{\micro\volt} absolute drift) over the course of \qty{49}{\min}, indicating stable and reproducible tuning of the QWC.

\section{Measurements of vapour-phase solvents}

\begin{figure*}[!htb]
    \centering
    \includegraphics[width=1\textwidth]{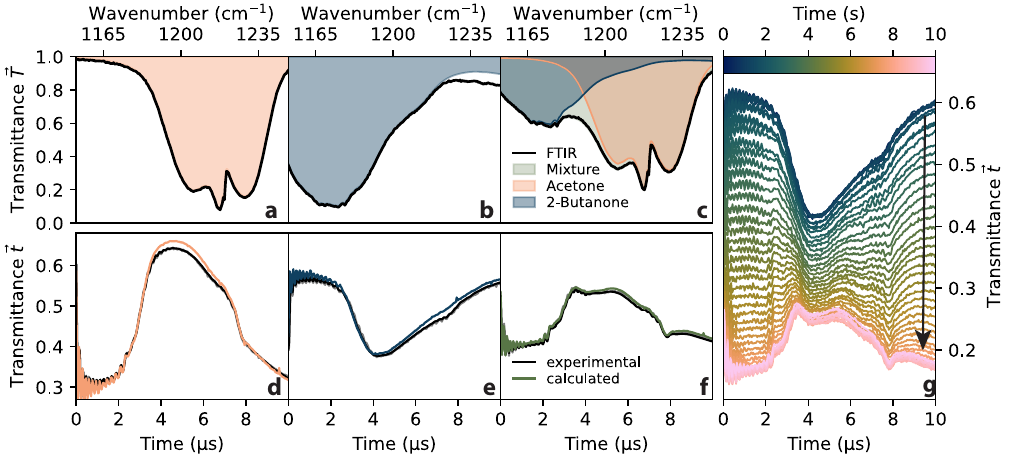}
    \caption{\textbf{Spectral and temporal transmittance measurements.} Acetone \textbf{(a)}, 2-butanone \textbf{(b)}, as well as a mixture of both \textbf{(c)}, were measured with an FTIR (black) with references fitted from the PNNL database (shaded area). In all instances \textbf{(d,e,f)}, the modulated QWC measurements demonstrate characteristic features consistent with the transmittances calculated from the established behaviour of the frequency comb (coloured lines). The results include both a single shot measurement of \qty{10}{\micro\second} (grey) and an average measurement of \qty{1}{\s} (black). For a dynamic measurement, 2-butanone is purged from the gas cell using a \qty{320}{\sccm} flux of diluted acetone, while the temporal transmittance $\vec{t}$ is continuously monitored \textbf{(g)}.}
    \label{fig: measurements}
\end{figure*}

Spectroscopic measurements were performed on nitrogen-diluted acetone and 2-butanone in their vapour phase, confined within a gas cell with a path-length of \qty{5}{\cm}. A transmittance reference $\vec{T}_\text{ref}$ was established by analysing the gases using a FTIR instrument equipped with a globar source. The concentrations were subsequently derived by fitting the obtained measurement data (depicted in black in Figure \ref{fig: measurements}a,b,c) with published reference spectra from the Pacific Northwest National Laboratory (PNNL) (illustrated by the shaded area in Figure \ref{fig: measurements}a,b,c, fit results in table \ref{tab:concentrations}) \cite{sharpe_gas-phase_2004}.

For data acquisition using the tunable QWC, a photovoltaic MCT sensor (Kolmar Technologies) was interfaced with an oscilloscope, while the QWC was swept through its resonance. Consequently, the observed time-dependent intensity trace $\vec{b}$ can be understood as the product of the laser's injection map $\mathbf{I}$ (the transpose of Fig. \ref{fig: comb tuning}a) and the wavenumber-dependent transmittance vector $\vec{T}$ of the gas under investigation (see Fig. \ref{fig: overview}):

\begin{equation}
  \mathbf{I} \cdot \vec{T} = \vec{b}, \quad \text{where}
  \begin{cases}
    \mathbf{I} & \text{is the intensity matrix,} \\
    \vec{T} & \text{is the transmittance vector,} \\
    \vec{b} & \text{is the detector signal.}
  \end{cases}
  \label{eq:matrix-multiplication}
\end{equation}

Or written more explicitly

\begin{equation}
    \begin{bmatrix} 
        \mathbf{I(f_1, \nu_1)}, & \cdots, & \mathbf{I(f_1, \nu_m)} \\ 
        \vdots & \ddots & \vdots \\ 
        \mathbf{I(f_n, \nu_1)}, & \cdots, & \mathbf{I(f_n, \nu_m)} 
    \end{bmatrix}
    \cdot
    \begin{bmatrix} 
        \mathbf{T_1}\\ 
        \vdots \\ 
        \mathbf{T_m} 
    \end{bmatrix}
    =
    \begin{bmatrix} 
        \mathbf{b_1}\\ 
        \vdots \\ 
        \mathbf{b_n} 
    \end{bmatrix}
\end{equation}

where each row of the matrix $\mathbf{I}$ represents a spectrum of the comb at a given injection frequency $f_i$ as a function of wavenumber $\nu_j$, $T_j$ is the wavenumber-dependent transmittance, and $b_i = \sum_{j=1}^m I(f_i,\nu_j)T_j$ is the spectrally integrated intensity measured as a voltage on the detector as a function of injection frequency (or equivalently, time). 

The voltage measured from the MCT following the passage of the laser through the previously mentioned samples $\vec{b}_{\text{sample}}$ can be transformed to a temporally resolved transmittance according to

\begin{equation}
    \vec{t}_\text{experimental} = \frac{\vec{b}_{\text{sample}} - \vec{b}_{\text{dark}}}{\vec{b}_{\text{nitrogen}}-\vec{b}_{\text{dark}}}
    \label{eq: temporal transmittance}
\end{equation}

where $\vec{b}_{\text{nitrogen}}$ denotes the trace obtained in the presence of only nitrogen and $\vec{b}_{\text{dark}}$ represents the dark reference with the laser beam blocked. Note that the dark background only has to be taken into account for the DC-coupled tunable QWC measurements and does not impact the AC-coupled FTIR measurements. Due to slow drifts in the alignment of the laser and detector temperature, $\vec{b}_\text{nitrogen}$ evolves slowly in time. These drifts are accounted for in the data processing. Note that we use $\vec{T}$ to refer to spectral transmittances, while $\vec{t}$ is used for temporal transmittances. Assuming that the sample concentration remains constant, both the detector output and temporal transmittance demonstrate periodic behaviour with a period of \qty{10.2}{\micro\second}, which facilitates signal averaging. Figure \ref{fig: measurements}abc presents \qty{10}{\micro\second} individual shot measurements (illustrated in grey) alongside \qty{1}{s} averaged results (depicted in black) for the three static samples under investigation. It is noteworthy that no averaging is needed to achieve sufficient data quality. The Allan deviations indicate drifts in the detector voltage starting from around \qty{10}{\milli\second} that are not apparent in the laser voltage's behaviour (reported in the supplementary material). This suggests that alignment shifts, or heating of the detector, rather than laser stability, are constraining the system's long-term stability.

Using equation \ref{eq:matrix-multiplication}, the temporal transmittance $\vec{t}$ can be calculated using the known behaviour of the laser $\mathbf{I}$ and the spectral transmittances $\vec{T}$ according to

\begin{equation}
    \vec{t} 
    = \frac{\mathbf{I} \cdot \vec{T}}{\mathbf{I} \cdot \vec{1}}
    = \mathbf{I}_\text{norm} \cdot \vec{T}
    \label{eq:matrix multiplication (norm.)}
\end{equation}

assuming that the measurements obtained using the tunable QWC are not subject to additional transmission attributed to the instrument itself, that is, $\vec{T}_{\text{FTIR}}=\vec{T}_\text{FM}$. Under this assumption, the normalised intensity matrix $\mathbf{I}_\text{norm}$ can be found by multiplying the measured matrix $\mathbf{I}$ by a diagonal matrix containing its row-wise sum from the left. Using the FTIR references $\vec{T}=\vec{T}_\text{ref}$, the expected temporal transmittances can be determined and presented alongside the measurement data (black in Fig. \ref{fig: measurements} d, e, f). These data reveal analogous functional forms and oscillatory behaviour during initial time periods. 

In addition to these measurements of static samples, to evaluate the speed of the proposed spectroscopic technique, the gas cell was filled with 2-butanone and subsequently flushed rapidly with a \qty{320}{\sccm} flow of nitrogen-diluted acetone. Temporal transmittance was recorded throughout the gas exchange using the tunable QWC (Fig. \ref{fig: measurements}g). Due to the insufficient temporal resolution provided by the FTIR, no reference measurement was obtained for this transient.

The obtained temporal transmittances $\vec{t}$ (Fig. \ref{fig: measurements}d,e,f) can be qualitatively understood by studying the tuning behaviour of the QWC, as illustrated in Fig. \ref{fig: comb tuning}a. At early times, characterised by significant negative detunings from resonance, the comb bandwidth is limited, approximately extending across the range \qtyrange{1190}{1220}{\per\cm}. This results in substantial overlap with the absorbance of acetone centred around \qty{1215}{\per\cm}, while remaining outside of the primary absorption feature of 2-butanone falling at \qty{1180}{\per\cm}. As a result, in these early times, the transmittance $\vec{t}$ is significantly lower in the acetone sample compared to the 2-butanone sample, despite their comparable peak absorbances (Fig. \ref{fig: measurements}). As the laser approaches its resonance after \qty{5}{\micro\second}, the QWC expands, roughly spanning the range \qtyrange{1160}{1245}{\per\cm}. The redistribution of most optical intensity from the free-running emission wavelength to the outer regions of the optical spectrum causes a substantial increase in transmittance $\vec{t}$ in acetone measurements, whereas it decreases for 2-butanone. The asymmetric nature of the spectral broadening in response to positive and negative detunings leads, for instance, to the asymmetry observed in the 2-butanone trace, which is crucial for mitigating the degeneracy of the matrix $\textbf{I}$.

\section{Reconstruction of spectral transmittances}

Depending on whether there is prior knowledge of the identity of the present gas species, two vastly different approaches to the reconstruction of the spectral transmission $\vec{T}$ from the temporal transmittance $\vec{t}$ can be implemented. 

\begin{figure*}[!hb]
    \centering
    \includegraphics[width=1\textwidth]{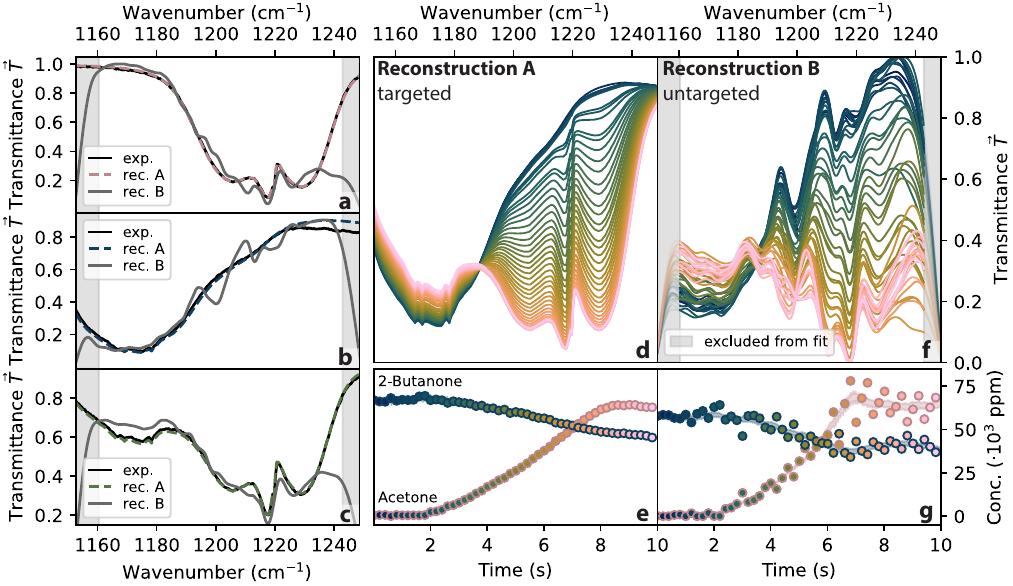}
    \caption{\textbf{Reconstruction of spectral transmittance}. The spectral transmittances $\vec{T}$ of acetone (\textbf{a}), 2-butanone (\textbf{b}), and the mixture of the two (\textbf{c}) are reconstructed to from the single-shot static measurements (black line) using the target reconstruction A (dotted, coloured line) and untargeted reconstruction B (grey, solid line). The fitted concentrations are reported in Table \ref{tab:concentrations}. The dynamic measurements covering the rapid replacement of 2-butanone with acetone vapour are reconstructed using the same targeted (\textbf{d}) and untargeted (\textbf{f}) approaches, with the obtained concentrations of both gases over time reported in the bottom figures (\textbf{e}, \textbf{g}). The region of high regularisation error of approach B (grey shaded region in \textbf{a},\textbf{b},\textbf{c},\textbf{f}) is excluded from the fit of the concentration.}
    \label{fig: reconstruction}
\end{figure*}

If the species are known, equation \ref{eq:matrix multiplication (norm.)} can be used as a basis of a fit function that determines $\vec{T}$ as the sum of known transmittances $\vec{T}_i$ with respective concentrations $c_i$ acting as fit parameters. According to the Beer-Lambert law, the total transmittance can be written as:

\begin{equation}
    \vec{T}_\text{fit} = \prod_{i=1}^N \vec{T}_i^{c_i} 
    = 10^{-\sum_i c_i L \vec{\alpha}_i}
\end{equation}

where $L = $\qty{5}{\cm} is the path length of the gas cell, $N$ is the number of species known to be present in the sample and $\vec{\alpha_i}$ denotes their respective absorption coefficient as function of wavenumber $\nu$. This approach, here referred to as targeted reconstruction \textbf{A}, can be straightforwardly implemented by minimising the residuals ($\mathbf{I}_\text{norm} \cdot \vec{T}_\text{fit} - \vec{t}_\text{experimental}$) using a least-squares algorithm. Assuming that either acetone (Fig. \ref{fig: reconstruction}a), 2-butanone (Fig. \ref{fig: reconstruction}b) are both (Fig. \ref{fig: reconstruction}c,d) present, the fits were performed on \qty{10}{\micro\second} single-shot data. The concentrations obtained for acetone are consistent with the FITR measurement ($\vert z \vert = 1.8 \sigma$), showing larger deviations for the other solvents ($3.6\sigma \leq \vert z \vert \leq8.3 \sigma$)  (as shown in table \ref{tab:concentrations}), with errors partially due to the offset in the FM measurements previously discussed. The reported uncertainties solely reflect the uncertainty inherent to the fit, and, notably, are comparable for both concentration assessments, whether derived from fitting the PNNL spectra to the FTIR or tunable QWC traces.

\begin{table}[ht]
  \renewcommand{\arraystretch}{2}
  \centering
  \resizebox{0.95\columnwidth}{!}{
  \begin{tabular}{|c|c|c|c|}
    \hline
     & \textbf{FTIR}    & \textbf{Reconstruction A} & \textbf{Reconstruction B} \\ \hline
    \textbf{Acetone}     & \num{54740+-90}   & \num{55080+-170}   & \num{60000+-1800}   \\ \hline
    \textbf{2-Butanone}  & \num{77200+-600}   & \num{79480+-230}   & \num{78300+-1200}   \\ \hline
    \textbf{Mixture} (ace.)    & \num{33860+-50}  & \num{33350+-60}  & \num{38800+-900}  \\
    (but.)                     & \num{16650+-40}  & \num{17320+-70}  & \num{13500+-600} \\
    \hline
  \end{tabular}
  }
  \caption{Concentrations of sample gases (in \unit{\ppm}) fitted from FTIR \\ and FM QWC transmittance measurements.}
  \label{tab:concentrations}
\end{table}

In order to further assess the precision and detection limits of tunable QWC spectroscopy, an additional dynamic measurement was performed in which the acetone-filled gas cell was rapidly flushed with \qty{2000}{\sccm} nitrogen flux (reported in the Supplementary). At each time step, the previously described reconstruction method \textbf{A} was executed and the acetone concentration was determined, demonstrating an exponential decrease in concentration as acetone was displaced by nitrogen. This process was accompanied by fluctuations in concentration, which could be attributable to turbulence within the chamber due to rapid gas exchange. The uncertainty of the fit was determined at each point of the reconstruction, indicating a largely constant relative uncertainty of \qty{0.45}{\%} while the acetone concentration remained above \qty{1000}{\ppm} and an absolute uncertainty of \qty{7.5}{ppm} below this threshold, including the zero-gas region. This observation suggests that for the current implementation, the limit of detection of acetone, ensuring a certainty of $3\sigma$, is \qty{23}{\ppm}.

The second approach to infer the spectral transmittances $\vec{T}$, referred to as untargeted reconstruction \textbf{B}, assumes no prior knowledge of the gas species present. Instead, equation \ref{eq:matrix multiplication (norm.)} is solved for $\vec{T}$ by finding the Moore-Penrose pseudo-inverse of $\mathbf{I}_\text{norm}$ and multiplying from the left, such that

\begin{equation}
    \vec{T}
    \approx \mathbf{I}_\text{norm}^{-1} \mathbf{I}_\text{norm} \cdot \vec{T}
    = \mathbf{I}_\text{norm}^{-1} \vec{t}. 
    \label{eq:matrix inversion}
\end{equation}

However, owing to the poor conditioning of the matrix, which retains a condition number of \num{122} even after the truncation of \qty{46}{\%} singular values, the inversion problem remains prone to numerical errors. In order to facilitate some degree of reconstruction despite these challenges, the matrix is subjected to Tikhonov regularisation, employing the first- and second-order discrete derivatives as the Tikhonov matrices. Furthermore, physical constraints are imposed on the fit by limiting the spectral transmittance $\vec{T}$ between 0 and 1. At the boundaries of the wavenumber region under examination, considerable regularisation errors are observed (transparent region in Fig. \ref{fig: reconstruction}f).

This approach permits the reconstruction to effectively capture the peak positions of both 2-butanone and acetone (Fig. \ref{fig: reconstruction}a,b,c,f ), and supports the subsequent fitting of concentrations using the PNNL reference spectra (Fig. \ref{fig: reconstruction}g). Reconstruction of the dynamic data accurately reveals overall trends, such as the increase in transmittance between \qtyrange{1160}{1180}{\per\cm} or the decrease in transmittance between \qtyrange{1205}{1235}{\per\cm}, with the prominent characteristic of acetone accurately identified at \qty{1217}{\per\cm} (Fig. \ref{fig: reconstruction}). The concentrations derived by fitting the PNNL spectra to these spectral transmittances, masking regions with high regularisation errors (transparent in Fig. \ref{fig: reconstruction}abcf), demonstrate the same behaviour as observed for reconstruction \textbf{A}, albeit with reduced precision and accuracy (Fig. \ref{fig: reconstruction}g). 


\section{Discussion and Outlook}

We have presented a fast and compact spectrometer, leveraging the reproducible tunability of the spectral envelope of a Quantum Walk Comb. The liquid-like nature of its intracavity light, stabilised by fast gain dynamics of the QCL, allows for reliable and fast tuning across an extensive spectral range in the mid-infrared spectra region via frequency modulation of the supplied RF signal. This enables instantaneous real-time transmittance measurement with an acquisition time of \qty{10}{\micro\second} across a wide bandwidth of \qty{97}{\per\cm}, employing a notably simplified optical configuration that eliminates moving components or matched comb pairs and comprises solely a driven comb source, the sampling region, and a detector. The temporal concentrations of gas species under examination can be ascertained through either targeted or untargeted analysis, depending upon the prior knowledge of the existing species and whether the emphasis lies on the precise reconstruction of dynamic concentrations or species identification. The system level of precision comparable to that attained by FTIR spectroscopy. The accuracy could be enhanced by improving the ruggedness of the system, for instance, by integrating the source, sampling region, and detector on-chip. More sophisticated modulation schemes, such as dual-tone injection \cite{piciocchi_frequency_2025}, are expected to enhance the numerical robustness of the reconstruction. Future research should encompass demonstrations of transient processes on the \qty{100}{\micro\second} timescale to evaluate the bounds of time-resolution, precision, and accuracy. The potential for on-chip integration could facilitate the development of robust and minituarised spectrometers for the investigation of molecular kinetics.


\printbibliography

\end{document}